\documentstyle[aps]{revtex}
\begin{document}
\tightenlines
\draft

\newcommand{\be}{\begin{eqnarray}}
\newcommand{\ee}{\end{eqnarray}}
\newcommand{\dia}{\!\!\!\!\!\not\,\,\,}

\twocolumn[\hsize\textwidth\columnwidth\hsize\csname
@twocolumnfalse\endcsname

\title{Electroweak baryogenesis in the standard model with strong
       hypermagnetic fields} 
\author{Alejandro Ayala and Gabriel Pallares}
\address{Instituto de Ciencias Nucleares\\
         Universidad Nacional Aut\'onoma de M\'exico\\
         Apartado Postal 70-543, M\'exico Distrito Federal 04510, M\'exico.}
\maketitle
\begin{abstract}

We show that in the presence of large scale primordial hypermagnetic  
fields, it is possible to generate a large amount of $CP$ violation 
to explain the baryon to entropy ratio during the electroweak phase
transition within the standard model. The mechanism responsible for
the existence of a $CP$ violating asymmetry is the chiral nature of the
fermion coupling to the background field in the symmetric phase which
can be used to construct two-fermion interference processes in analogy to
the Bohm-Aharanov effect. We estimate that for strong hypermagnetic
fields $B_Y=(0.3-0.5)\ T^2$ the baryon to entropy ratio can be
$\rho_B/s=(3 - 6)\times 10^{-11}$ for slowly expanding bubble walls.

\end{abstract}
\pacs{PACS numbers: 12.15.--y, 98.80.Cq }
\vskip2pc]

The understanding of the mechanism that generates the observed excess
of baryons over antibaryons in the universe represents one of the most
challenging problems for particle physics as applied to cosmology. Any theory
that aims to explain such excess has to meet the three well know Sakharov
conditions\cite{Sakharov}, namely: (1) Existence of interactions that
violate baryon number; (2) {\it C} and {\it CP} violation; (3)
departure from thermal equilibrium. It is well known that the
above conditions are met in the standard model (SM) of electroweak interactions
if the electroweak phase transition (EWPT) is of first order. This raised the
interesting possibility that the cosmological phase transition responsible for
particle mass generation --that took place at temperatures of order
$100$ GeV-- could also explain the generation of baryon
number. Consequently, a great deal of effort was devoted to explore
such a possibility~\cite{Trodden}. Nowadays, the consensus however is
that the minimal SM as such, cannot explain the observed
baryon number. The reason is that the EWPT turns out to be only too
weakly first order which in turn implies that any
baryon asymmetry generated at the phase transition was erased by
the same mechanism that produced it, {\it i.e.} sphaleron induced 
transitions~\cite{Kajantie}. Moreover, the amount of $CP$ violation
coming from the CKM matrix alone cannot account by itself for the
observed asymmetry, given that its effect shows up in the coupling of the
Higgs with fermions at a high perturbative order~\cite{Dine}, giving
rise to a baryon to entropy ratio at least ten orders of magnitude
smaller than the observed one.

Nevertheless, it has been recently pointed out that, provided
enough {\it CP} violation exists, the above scenario
could significantly change in the presence of large scale primordial
magnetic fields~\cite{{Giovannini},{Elmfors}} (see however
Ref.~\cite{Skalozub}), which can be 
responsible for a stronger first order EWPT. The situation is similar
to a type I superconductor where the presence of an external magnetic field
modifies the nature of the superconducting phase transition due to the
Meissner effect. 

Though the origin and nature of these primordial fields is a subject
of current research, its existence prior to the EWPT epoch can certainly
not be ruled out~\cite{reviews}. The only constraint on their strength
comes from considerations on the measured anisotropies of the cosmic 
microwave background and on nucleosynthesis calculations~\cite{Durrer}. 

Recall that for temperatures above the EWPT, the SU(2)$\times$U(1)$_Y$
symmetry is restored and the propagating, non-screened vector modes that
represent a magnetic field correspond to the U(1)$_Y$ hypercharge
group. Thus, in the unbroken phase, any primordial magnetic fields 
belong to the hypercharge group instead of to the U(1)$_{em}$ group
and are therefore properly called {\it hypermagnetic} fields.

In this paper we show that the existence of such primordial hypermagnetic
fields also provides a mechanism to produce a large enough amount
of {\it CP} violation during the EWPT to possibly explain the observed 
baryon to entropy ratio in the SM. This can happen during the
reflection of fermions off the true vacuum bubbles nucleated during the
phase transition through an interference process equivalent to the 
Bohm--Aharanov effect, given that in the unbroken phase, fermions
couple chirally to hypermagnetic fields with the hypercharge. The
chiral nature of this coupling implies that it is possible to build a
{\it CP} violating asymmetry dissociated from non-conserving baryon number
processes that can then be converted to baryon number in the unbroken
phase where sphaleron induced transitions are taking place with a large
rate. The existence of such asymmetry provides a bias for baryon over
antibaryon production. In the absence of primordial magnetic 
fields, this mechanism has been proposed and studied in
Refs.~\cite{{Dine},{Cohen},{Nelson}} in extensions of the SM. 

To describe the EWPT, we start by writing the effective, finite
temperature Higgs potential which, including all the one-loop effects
and ring diagrams, looks like
\be
   V_{\mbox{eff}}\ (\phi,T)=\frac{\gamma}{2}(T^2 - T_c^2)\ \phi^2
                  -\delta\ T\ \phi^3 +\frac{\lambda}{4}\ \phi^4\, ,
   \label{effectivepot}
\ee
where $\phi=\sqrt{2}(\Phi^\dagger\Phi)^{1/2}$ is the strength of the
SU(2) Higgs doublet $\Phi$ whose vacuum expectation value $v$ is given by
\be
   \langle\Phi\rangle = \frac{v}{\sqrt{2}}\, .
   \label{expectation}
\ee
The parameters $\gamma$, $\delta$ and $\lambda$ have been computed
perturbatively to one loop and can be expressed in terms of $v$,
the SU(2) gauge boson masses and the top mass. Their explicit
expressions can be found elsewhere (see for example
Ref.~\cite{Elmfors}). $\delta$ is the parameter responsible for the first
order nature of the phase transition. It is the parameter that gets
enhanced in the presence of hypermagnetic fields. $T_c$ is the
critical temperature at which spinodal decomposition proceeds.

We can write the effective potential in a more transparent form~\cite{Turok} 
by introducing the dimensionless temperature $\vartheta$ and the
dimensionless Higgs field strength $\varphi$ 
\be
   \vartheta &=& \frac{\lambda\ \gamma}{\delta^2} 
   \left[ 1 - \left(\frac{T_c}{T}\right)^2\right]\, \nonumber\\
   \varphi &=& \frac{\lambda}{\delta\ T}\phi\, ,
   \label{effs}
\ee
in terms of which, the effective potential, Eq.~(\ref{effectivepot}),
becomes
\be
   V_{\mbox{eff}}\ (\varphi) = \delta\ T
   \left(\frac{\delta\ T}{\lambda}\right)^3
   \left(\frac{\vartheta}{2}\varphi^2 - \varphi^3 + \frac{1}{4}\varphi^4
   \right)\, .
   \label{effV}
\ee
For simplicity, we work in the approximation where the energy
densities of both the unbroken and broken phases are degenerate. This
happens for a value of $\vartheta=2$. In this approximation, the phase
transition is described by a one-dimensional solution for the Higgs
field, called the {\it kink}, which separates the two phases. This is
given by
\be
   \varphi (x) = 1 + \tanh (x)\, ,
   \label{kink}
\ee
where the dimensionless position coordinate $x$ is 
\be
   x = \frac{\delta\ T}{\sqrt{2\lambda}}\ r\, .
   \label{dimensionlessr}
\ee
The parameter $\sqrt{2\lambda}/(\delta\ T)$ represents the width of
the domain wall~\cite{Liu}. It can also be checked that this parameter 
becomes smaller in the presence of hypermagnetic fields. 

In terms of the kink solution we can see that $x=-\infty$ represents
the region outside the bubble, that is the region in the symmetric
phase. Conversely, for $x=+\infty$, the system is inside the bubble,
that is in the broken phase. The kink wall propagates with a velocity
determined by its interactions with the surrounding plasma. This
velocity can be anywhere between 0.1--0.9 the speed of
light~\cite{Megevand}.

Given the above background scalar field, the problem of fermion
reflection and transmission through the domain wall in the presence of
an external magnetic field can be cast in terms 
of solving the Dirac equation. In the wall's rest frame, this can be
written as 
\be
   \left( p\dia\ -\ a\ A\!\dia\!\ -\ 
   \frac{\delta\ T}{\sqrt{2\lambda}}\ 
   \xi\ \varphi (x)\right)\Psi ({\mbox x})=0\, ,
   \label{dirac}
\ee
where $\xi=2m/m_H$, $m$ and $m_H$ being the fermion and Higgs masses
at zero temperature, respectively and $\Psi$ the fermion wave
function. Also, $A^{\mu}=(0,{\mathbf A})$ is the four-vector potential
related to the external magnetic field by  
${\mathbf B}=\nabla\times {\mathbf A}$. In the symmetric phase,
${\mathbf A}={\mathbf A}_Y$ corresponds to the hypermagnetic field
vector potential whereas in the broken phase, where only the Maxwell 
projection is unscreened~\cite{{Giovannini},{Elmfors}}, 
${\mathbf A}={\mathbf A}_{em}$ corresponds to the ordinary photon
vector potential. $a$ is the fermion coupling to the external field;
$a=g'{\mathsf Y}/2$ in the symmetric phase with ${\mathsf Y}$
the fermion's hypercharge and $g'$ the U(1)$_Y$ coupling constant; 
$a=e$, the electric charge, in the broken phase. 

In the absence of a background magnetic field, the solution to
Eq.~(\ref{dirac}) when considering the bubble wall as a planar
interface has been found in Ref.~\cite{Ayala}. Let
$\tilde\Psi_h ({\mbox x})$ be a solution with a definite helicity
($h=L,R$) to Eq.~(\ref{dirac}) in the absence of an external magnetic 
field --since we will be interested in fermion solutions in the symmetric
phase where particles are massless, helicity is a good quantum
number. It can easily be shown that in the presence of a magnetic
field, the solution to Eq.~(\ref{dirac}) with definite helicity is
given by
\be
   \Psi_h ({\mbox x}) = \tilde\Psi_h ({\mbox x})
   e^{-i\frac{g'}{2}{\mathsf Y}_h\int^{\mbox{\small{x}}}
   {\mathbf A}_Y\cdot d{\mathbf x}'}
   \label{solhel}
\ee
where ${\mathsf Y}_h$ is the hypercharge of the corresponding helicity
mode. 

Since the presence of the hypermagnetic field shows up as a phase
in the fermion wave function, it follows that its effect will dissapear when 
considering a process that involves the square of a single fermion wave
function. However, the effect will be revealed during an interference
process. Since this requires that at least two particles be present
within a spatial volume of order $\lambda^3$ per unit time, where
$\lambda$ is the de Broglie fermion wave length, we expect that these
processes are supressed with respect to those involving a single
particle. We will come back later to estimate the probability for such
processes. For the time being, let us consider a situation in which the
fermion amplitude at a space point a distance $l$ away from the the
bubble wall receives contributions from fermions coming from the
symmetric phase which are reflected from two points on the bubble wall
separated by a distance $d$. For simplicity, we take $d\ll l$, in this 
case, we can consider that the fermion propagation is approximately
directed along the direction perpendicular to the bubble wall. Since
the coupling of the given helicity mode with the hypermagnetic field
is chiral, the square of the amplitude representing the interference
of left-handed fermions $|M_L|^2$ will differ from the square of the
amplitude for right-handed interfering antifermions $|\bar{M}_R|^2$ and
consequently, an axial asymmetry ${\mathcal A}$ is built in front of
the bubble wall. Within the above approximations, it is easy to show
that the explcit expression for the axial, $CP$ violating asymmetry 
is given by
\be
   {\mathcal A}&\equiv&|M_L|^2 -\ |\bar{M}_R|^2\nonumber\\
               &=&{\mathcal R}\sin (\Upsilon_+/2)
   \sin (\Upsilon_-/2)\, ,
   \label{asymmetry}
\ee
where 
\be
   \Upsilon_\pm=\frac{g'}{2}\left({\mathsf Y}_R\pm
   {\mathsf Y}_L\right)
   \Theta\, .
   \label{upsi}
\ee
${\mathcal R}$ is the fermion reflection coefficient~\cite{Ayala}, 
common to left and right-handed modes and depends on the parameter 
$\xi$ and the fermion energy, being equal to $1$ for energies below
twice the zero temperature mass of the fermion. $\Theta$ is the
hypermagnetic field flux through the area defined by the two
trajectories of the interfering fermions. Taking the case of a
constant hypermagnetic field perpendicular to this area, 
$\Theta=B_Yld/2$, with $B_Y=|{\mathbf B_Y}|$. 
Form
Eq.~(\ref{asymmetry}) we see that the asymmetry disappears for a
vanishing hypermagnetic field.     

We now follow Ref.~\cite{Nelson} and construct the net axial charge
flux ${\mathcal F}$ in front of the expanding bubble wall in the symmetric
phase. This flux receives contributions both from the reflected
fermions within the symmetric phase and from those fermions transmitted
from the bubble's interior. From $CPT$ conservation, fermion
transmission from the broken to the symmetric phase is related to
fermion reflection in the symmetric phase. Therefore, the explicit
expression for ${\mathcal F}$ is given by~\cite{Torrente}
\be
   {\mathcal F}&=&\frac{1}{2\pi^2\gamma}\int_0^\infty dp_l
   \int_0^\infty dp_t\ p_t\nonumber\\
   &&\left[\ f^s(-p_l,p_t) - f^b(p_l,p_t)\ \right]\ {\mathcal A}\, ,
   \label{chargeflux}
\ee 
where the bubble's wall velocity in the fluid frame is $u$ and
$\gamma=1/\sqrt{1-u^2}$ is the Lorentz factor. $p_l$, $p_t$
are the longitudinal and transverse components of the fermion's
momentum and $f^s$, $f^b$ are the fermion equilibrium fluxes
(neglecting Fermi blocking factors) in the symmetric and broken
phases, respectively

\vspace{0.2cm} 
\( \begin{array}{cl}
   f^s=\frac{p_l/E^s}{\exp [\gamma (E^s-up_l)/T]+1}, &
   E^s=\sqrt{p_l^2+p_t^2} \\
   f^b=\frac{p_l/E^b}{\exp [\gamma (E^b+up_l)/T]+1}, &
   E^b=\sqrt{p_l^2+p_t^2+m^2}\, .
\end{array} \) 
\vspace{0.2cm}

Since there is no net fermion number flux through the wall, the axial 
charge ${\mathcal F}$ and hypercharge ${\mathcal F}_Y$ fluxes are trivially 
related by ${\mathcal F}_Y={\mathcal F}/4$.
 
The rate of baryon number density production is given in terms of the
rate of baryon number violation per unit volume $\Gamma_B$ and the 
partial derivative of the free energy with respect to baryon number
$\partial F/\partial B$ by~\cite{Dine}
\be
   \dot{\!\tilde{\rho}}_B=-\frac{\Gamma_B}{T}
   \frac{\partial F}{\partial B}\, .
   \label{rate} 
\ee
The quantity $\partial F/\partial B$ corresponds to the baryon number
chemical potential $\mu_B$ and represents the force pushing the
universe towards its equilibrium baryon number value. Only those
processes which happen fast enough with respect to the time that the
reflected fermions in the symmetric phase (or those that passed from the
broken to the symmetric phase) spend before being retaken by the
expanding wall, will contribute to drive baryon number towards
equilibrium. This time is called the {\it transport time} $\tau$ and 
estimates show~\cite{Nelson} that it is of order $\tau\sim 100/T$. 
Since the only other interactions, besides SU(3)$\times$U(1) gauge
boson exchange and family-diagonal SU(2) gauge boson exchange, with a
large enough rate are the top quark Yukawa interactions (given the
large top Yukawa coupling), then the axial asymmetry must reside
basically in the form of top quarks. Enforcing this condition together
with that of having initially zero net baryon and lepton numbers one
finds~\cite{Nelson} 
\be
   \mu_B=\frac{\partial F}{\partial B}=
   -\frac{4\rho_Y}{(1+2n)T^2}\, ,
   \label{chem}
\ee
where $\rho_Y$ is the hypercharge density, $n$ is the number of scalar 
doublets in the theory and the minus sign comes from the assumption
that the net axial asymmetry is in the form of a right-handed top
number. For the SM, $n=1$.  

It is now straightforward to integrate Eq.~(\ref{rate}) to get the net
baryon number density in terms of the axial flux, with the 
result~\cite{Nelson}
\be
   \tilde{\rho}_B=\frac{\Gamma_B}{3T^3}\frac{\tau}{u}{\mathcal F}\, ,
   \label{baryonnumb}
\ee
Equation~(\ref{baryonnumb}) has to be corrected by recalling that the
axial asymmetry, Eq.~(\ref{asymmetry}), is built from two-fermion
interfering processes. We estimate the probability $\eta$ of finding a 
second fermion in the trajectory of the first one by assuming that there is
no initial correlation between these two particles. The probability to
find the second one is thus given by the ratio of the number of
particles per unit area and time that bounce off the bubble wall
making an angle $\theta$ within a solid angle subtended by a linear 
dimension on the order of the particles wave length ($\lambda\sim T^{-1}$)
to the incoming flux. $\theta$ is such that
\be
   \cos\theta = \frac{l}{\sqrt{l^2 + (d/2)^2}}\, .
   \label{teta}
\ee 
Taking $p$, $dp\sim T$ and neglecting Fermi blocking factors, $\eta$
is given by 
\be
   \eta = \frac{2}{3\pi\zeta (3)}\frac{\lambda^2\ l\ d/2}
   {[\ l^2 + (d/2)^2\ ]^2}\frac{1}{1+e}\, ,
   \label{eta}
\ee 
where $\zeta$ is the Riemann zeta function and the last factor comes
from a simple Fermi-Dirac distribution evaluated at $E=T$. 

Notice that in the wall's rest frame, the interference point cannot be
further apart than the particle's mean free path 
$\lambda_{\mbox {mfp}}(\sim(1 - 10)/T\cite{Liu})$, since otherwise 
fermions will rethermalize and no net baryon number could be
produced. Accounting for this probability factor, the net baryon to
entropy ratio produced during the EWPT can be written as
\be
   \rho_B/s&=&\eta\tilde{\rho}_B/s\nonumber\\
   &=&\eta\frac{270\kappa\alpha_{\mbox w}^4}{12\pi^2 g_\star}
   \frac{\tau{\mathcal F}}{uT^2}\, ,
   \label{ratiobarent}
\ee
where we have used that for the EWPT epoch, the entropy density is
given by $s=2\pi^2g_\star T^3/45$ with the effective number of degrees 
of freedom $g_\star \simeq 107$ and that
the rate of baryon number violation per unit volume in the symmetric
phase is given by
\be
   \Gamma_B=3\kappa\alpha_{\mbox w}^4T^4\, ,
   \label{rateex}
\ee
with $\alpha_{\mbox w}=g^2/4\pi$, $g$ the SU(2) coupling constant and 
$\kappa\sim 1$. 

Equation~(\ref{ratiobarent}) is valid when the rate of
baryon number violation is much smaller than the rate at which the
fermions are recaught by the expanding wall, which in turn sets a lower
limit for the wall's velocity given by
\be
   3\kappa\alpha_{w}^4T\tau\ll u\, .
   \label{limit}
\ee
For definitiveness, we set $\kappa =1$, $m_h=T=100$ GeV and
$d=l/2$. (For the SU(2) and U(1)$_Y$ couplings, we use $g=0.637$
and $g'=0.344$, respectively~\cite{Carrington}). We find that for a
wall velocity $u=0.1c$  
\be
   \rho_B/s=(3 - 6)\times 10^{-11}
   \label{firstnum}
\ee
whereas for $u=0.6c$
\be
   \rho_B/s=(0.7 - 1.3)\times 10^{-11}\, ,
   \label{secnum}
\ee
for $l\sim 9/T$ and $B_Y=(0.3 - 0.5)\ T^2$. These are large but still 
not ruled out values for the background field strength~\cite{Durrer},
compatible with a Higgs mass of order 100 GeV and a phase transition
analog to a type I superconductor~\cite{Elmfors}. Recall that the
experimental value based on nucleosynthesis calculations~\cite{Kolb} is
\be
   \rho_B/s=(0.1 - 1.4)\times 10^{-10}\, ,
   \label{expnum}
\ee
which means that the estimates based on the present analysis are
within the experimental limits, at least for slowly expanding bubbles.

In conclusion, we have shown that in the presence of strong, large
scale primordial hypermagnetic fields, it is possible to generate a
large amount of $CP$ violation that combined with a stronger first
order EWPT --also produced by the hypermagnetic fields-- could account 
for the observed baryon number to entropy ratio within the SM. 
The fact that fermions couple chirally to background hypermagnetic 
fields in the symmetric phase makes it possible to build a $CP$
violating asymmetry by considering two fermion interfering processes in
an equivalent way to the Bohm-Aharanov effect. This asymmetry is
converted into baryon number by sphaleron induced processes in the
symmetric phase and preserved when these fermions are recaught by the
expanding bubble wall.  

Given that some of the parameters describing the dynamics of the EWPT
are not very precisely determined, it is clear that work on that
direction is necessary and the possibility that baryogenesis could be 
realized within the SM could certainly stimulate this kind of work.

Support for this work has been received in part by CONACyT-Mexico
under grant numbers 32279-E and 29273-E.


\begin{references}

\bibitem{Sakharov}
A. D. Sakharov, Pis'ma Zh. Eksp. Teor. Fiz. {\bf 5}, 32 (1967) [JETP
Lett. {\bf 5}, 24 (1967)].

\bibitem{Trodden}
For a recent review on the subject see M. Trodden, \rmp {\bf 71},
1463 (1999).

\bibitem{Kajantie}
K. Kajantie, M. Laine, K. Rummukainen and M. Shaposnikov,
Nucl. Phys. B {\bf 466}, 189 (1996).

\bibitem{Dine}
M. Dine, \lq\lq Baryogenesis: Electroweak and otherwise'', Proceedings
of the 1994 TASI summer school {\it CP violation and the limits of the
standard model}, 507-548. 

\bibitem{Giovannini}
M. Giovannini and M. E. Shaposhnikov, \prd {\bf 57}, 2186 (1998).

\bibitem{Elmfors}
P. Elmfors, K. Enqvist and K. Kainulainen, Phys. Lett. B {\bf 440},
269 (1998). 

\bibitem{Skalozub}
V. Skalozub and V. Demichik, \lq\lq Can baryogenesis survive in the
standard model due to strong hypermagnetic field?'', {\bf hep-ph/9909550}.

\bibitem{reviews}
For recent reviews on the origin, evolution and some cosmological 
consequences of primordial magnetic fields see: K. Enqvist,
Int. J. Mod. Phys. {\bf D}, 331 (1998); R. Maartens, \lq\lq Cosmological
magnetic fields'', International Conference on Gravitation and
Cosmology, India, Jan. 2000, {\bf astro-ph/0007352} and references therein.

\bibitem{Durrer} 
R. Durrer, P. G. Ferreira and T. Kahniashvili, \prd {\bf 61}, 043001
(2000); K. Jedamzik, V. Katalini\'c and A. V. Olinto, \prl {\bf 85},
700 (2000).

\bibitem{Dine}
M. Dine, O. Lechtenfield, B. Sakita, W. Fischel and J. Polchinski,
Nucl. Phys. B {\bf 342}, 381 (1990).

\bibitem{Cohen}
A. G. Cohen, D. B. Kaplan and A. E. Nelson, Phys. Lett. B {\bf 263},
86 (1991).

\bibitem{Nelson}
A. E. Nelson, D. B. Kaplan and A. G. Cohen, 
Nucl. Phys. B {\bf 373}, 453 (1992).

\bibitem{Turok}
N. Turok, \prl {\bf 68}, 1803 (1992).

\bibitem{Liu}
B. H. Liu, L. McLerran and N. Turok, \prd {\bf 46}, 2668 (1992).

\bibitem{Megevand}
For a recent discussion on the dynamics of EWPT, see A. Megevand,
\lq\lq Development of the electroweak phase transition and
baryogenesis'', {\bf hep-ph/0006177}.

\bibitem{Ayala}
A. Ayala, J. Jalilian-Marian, L. McLerran and A. P. Vischer, \prd 
{\bf 49}, 5559 (1994).

\bibitem{Torrente}
E. Torrente-Lujan, \prd {\bf 60}, 085003 (1999). 

\bibitem{Carrington}
M. E. Carrington, \prd {\bf 45}, 2933 (1992).

\bibitem{Kolb}
E. Kolb and M. Turner, \lq\lq The early universe'' (Addison-Wsley, New
York, 1990). 

\end{references}
\end{document}